# Achieving Network Slice Communication Service Distribution across 5G Micro-Operator Multi-tenants

Idris Badmus, Abdelquoddouss Laghrissi and Ari Pouttu
Centre for Wireless Communication
University of Oulu
Oulu, Finland
firstname.lastname@oulu.fi

*Abstract*— Network slicing is the means of logically isolating network capabilities from a "one size fits all" to a set of different slices where each slice is responsible for specific network requirements. In the same light, the micro-operator concept is targeted at local deployment of 5G for vertical specific services. When a network slice instance (NSI) is created across the access and core networks, it is important to understand how the communication service from the created slice instance will be distributed to different tenants and their end-users. In this vein, we propose a network slice communication service distribution technique for local 5G micro-operator deployment scenarios. This is achieved by expanding/leveraging the communication service management function (CSMF) defined by 3GPP into a multi-tenant manager and a communication services orchestrator. We further describe how the communication service orchestrator will address all the possible multitenant-slice situations that can exist during the distribution of a network slice instance to multiple tenants. The generic technique introduced in this paper will proffer a researchable solution for not only network slice communication service distribution across different micro-operator's tenants but pave the way to support future use cases especially when the allocated slice is responsible for multiple tenants or when a tenant requests multiple NSIs.

## I. Introduction

The idea of network slicing involves a sophisticated and careful creation of multiple end-to-end virtual network slices within a single network, whereby each created slice is responsible for different network requirements (i.e. eMBB, uRLLC and mMTC slices) [1], [2]. To achieve network slicing, it is important to leverage existing network softwarization concepts such as software defined networks and network function virtualization (SDN/NFV) [3], [4], while maintaining network resources usage across different administrative and technological domains [5].

Meanwhile, the micro-operator concept involves having a completely new local 5G network away from the Mobile Network Operator (MNO) network in terms of radio spectrum and infrastructure [6], [7]. The operations of the micro-operator networks are generally targeted at vertical specific services and due to the fact that different vertical tenants have different requirements, it is paramount to have a micro-operator network serving multiple tenants with separate network requirements. This can only be achieved with network slicing in a micro-operator network [8].

In order to describe an efficient micro-operator network model, authors in [9]–[11] have proposed different deployment scenarios of a micro-operator that can exist. The deployment scenarios include the Closed, Open and Mixed networks. Each of the deployment scenarios are described to cover all possible cases occurring whenever a stakeholder wants to have a micro-operator network. Thus, to achieve a complete end-to-end network slicing across these micro-operator deployment scenarios from the moment that a slice is requested to the moment when the slice is allocated, we need to understand not only how the NSI is created across the access and core networks, but also how the created NSIs will be distributed across different tenants based on their deployment scenarios or use cases.

Generally, to activate a network slice across tenant's end-users, 3GPP [1] introduced the CSMF as part of the management functionalities required for network slicing following the network slice management function (NSMF) and the network slice subnet management function (NSSMF). The NSSMF is responsible for managing and orchestrating different network function (NFs) to make up the network slice subnet instance (NSSI) or the network service (NS) and the NSMF is responsible for managing and orchestrating different NSSIs to form the NSI (i.e eMBB, uRLLC or mMTC slice instances). The CSMF on the other hand is responsible for aggregating and translating different slice requests as well as distributing the created NSIs to the required tenant's end users as a network slice communication service.

However, even though the implementation of the NSSMF and the NSMF have been achieved to some extent by specific open source bodies such as Open Source Mano (OSM) [12], the implementation of the CSMF for the distribution of created NSIs to tenant's end-users is still a matter that is yet to be solved. This is an extensive and vital research area especially in use cases where the NSI is responsible for multiple tenants or when a single tenant is requesting multiple NSIs.

In this paper, we propose a foundational technique by which the network slice communication service can be transmitted and distributed across multi-tenants, considering different micro-operator deployment scenarios. Our proposed technique leverages the CSMF capabilities by introducing a multi-tenant manager and a communication service orchestrator in the network slicing architecture of a micro-operator to cover all possible multitenant-slice situation types that can exist. While the multi-tenant manager handles the CSMF responsibility of aggregating and translating network slice requests from different tenants, the communication service orchestration is responsible for the distribution of the created slice request to different tenants based on the situation types. The situation types are basically cases that can exist whenever an established slice is to be distributed to different tenants. The ones covered in this paper are the following;

-*Situation1: One Network slice instance distributed to one tenant*
-*Situation2: One Network slice instance is distributed across multiple tenants*
-*Situation 3: Multiple slice instances are to be distributed to a single tenant*
-*Situation 4: Multiple slice instance are to be distributed to multiple tenants*

The implementation technique by which the communication service orchestration will use in achieving network slice distribution for the four situation types will be extensively described in the paper

The remaining part of this work is arranged as follows. Section II presents existing literatures on slice instance distribution. Previous research work on micro-operator network model and how network slicing can be achieved for each deployment scenario is presented in section III. The proposed network slice distribution technique across multi-tenants in a micro-operator network is presented in section IV. The Concluding section contains the general summary of the work ironed out so far and the possible future contributions.

## II. EXISTING LITERATURES ON SLICE INSTANCE DISTRIBUTION

The idea of having a network slice distributed across different tenants is an important branch of achieving end-to-end network slicing. Generally, whenever a virtualized slice instance is established across the access network (AN) and core network (CN), it is paramount to direct the communication service from the established slice to a tenant in a dynamic and efficient way throughout the lifecycle of the slice. It should be noted that the distribution of an established network slice is always prone to the use case that can exist and thus, a different use case may require a different approach for the slice instance distribution. To achieve multi-tenancy slice distribution and resource allocation, authors in [13] introduced the concept of 5G network slice broker to facilitate on-demand resource allocation for each tenant and perform admission control based on monitoring and forecasting. The 5G network slice broker resides with the network provider or the infrastructure provider, whereby all the required interfaces and functionalities are detailed. However, the approach in [13] introduces only the idea of a multi-tenancy in network slicing without considering how different NSIs will be transmitted across different tenants. To further support slice distribution across a multi-tenant situation, authors in [14] introduced the concept of having a multi-tenant NFV MANO that will be responsible for resources management and orchestration at the tenant level. The MANO as a Service (MANOasS) concept is proposed as an extension of the existing ETSI NFV-MANO model [15], by leveraging the virtualization abstraction of the ETSI NFV-MANO to the tenants layer. This is described as the Tenant MANO (t-MANO). t-MANO will ensure that tenants have the capability to manage and orchestrate their own slice instances provided by the network operator. However, this approach adds complication to the distribution of the slice instance by adding a set of managers and orchestrators at tenant level. Also, this approach is not aligned with the 3GPP standard of CSMF handling the slice distribution to different tenants and their end users. As stated earlier, 3GPP [1] introduced the CSMF functionality for the aggregation of slice requests for tenants and the distribution of slice instances as communication service to different end users. However, even though this functionality is introduced by 3GPP, the implementation technique to support different use cases for a multiple tenant scenario has not been addressed.

With regards to implementation techniques, OSM [12], have proposed a network slicing implementation across different technological and administrative domains. However, the current implementation covers only the NSMF and the NSSMF, two out of the three 3GPP defined network slicing functionalities [1]. OSM [12] uses a layered approach in achieving network slicing in a virtualized network environment. This is done such that the Virtual Machines (VMs) are the basis of the layers having an image instance in a Virtual Infrastructure Manager (VIM); multiple VMs are described together with the Virtual Network Function Descriptor (VNFD) to form the Virtual Network Functions (VNFs) as another layer; furthermore, different VNFs are described together with Network Service Descriptor (NSD) to form the NS or NSSI, managed by the NSSMF; and finally different NSSIs are described together with a Network Slice Template (NST) to form the NSI, managed by the NSMF. Thus, the current implementation technique covers the NSMF and NSSMF, without the Communication Service Management Function (CSMF), which is responsible for aggregating and translating slice requests and determining how the NSIs will be distributed as a communication service across different tenants and their end users.

Hence, our proposed approach consists of leveraging in the 3GPP's CSMF functionality by proposing an implementation technique to achieve slice request aggregation and slice distribution for different tenants by introducing a multi-tenant manager and a network slice orchestrator. Our design considers all possible situations types and covers any possible use case when a slice instance is to be distributed to multiple tenants.

## III. NETWORK SLICNG FOR A LOCAL 5G MICRO-OPERATOR NETWORK MODEL

In light of achieving network slicing for a micro-operator, the micro-operator network model in Fig.1 has been proposed [7], [16] in order to accommodate various verticals and their tenants, whereas each tenant contains a different set of end users.

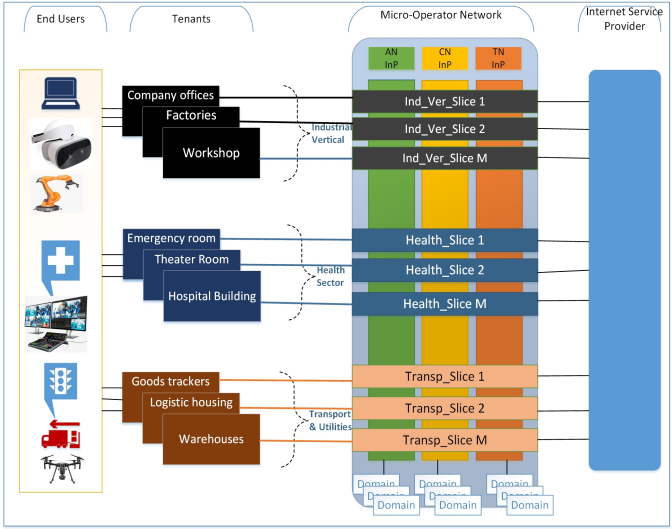

Fig 1. Micro-operator network model.

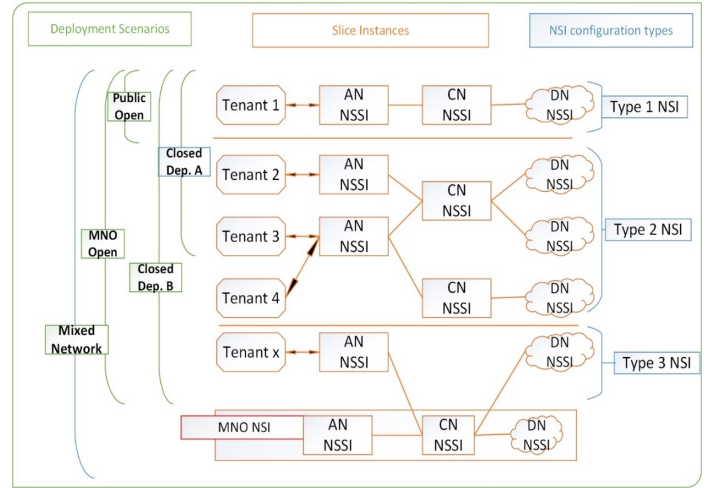

Fig 2. NSI configuration types for micro-operator deployment scenarios [8].

Based on the network model in Fig. 1, a micro-operator network will be responsible for serving multiple vertical's network requirements. Each vertical is made up with different tenants that will be allocated a network slice based on the service requirements of the tenant's end users. As highlighted in the previous section, A micro-operator network can be Closed, Open or Mixed, and the description for each deployment scenario can be seen in Table I.

Generally, Network slicing in a micro-operator is categorized based on the **NSI Configuration Type** that is peculiar to each deployment scenario [11]. The NSI configuration type describes how the RAN and CN NSSIs will be shared to fulfill specific network services. Three NSI configuration types are proposed in [11]. These include Type 1, Type 2 and Type 3 NSI configurations. Type 1 NSI configuration is targeted at tenants with very strict latency requirement and hence there won't be any shared subnets between the dedicated slice NSSI and other NSSIs in the network. Type 2 NSI configuration is targeted at tenants with less strict latency requirement and as such there can be shared subnets with other NSSIs of the same network. Type 3 NSI configuration happens when there is need to have shared NSSIs with the external network. Fig. 2 shows how each micro-operator network deployment scenario will be mapped to the respective NSI configuration type.

IV. NETWORK SLICE DISTRIBUTION TECHNIQUE FOR MULTI-TENANTS IN A MICRO-OPERATOR NETWORK

In this section, we present a new framework that will support four situation types that can exist for any use case during a network slice communication service distribution across a multi-tenant network. Basically, after a network slice instances is established, the distribution of the communication service across different available tenants in the network can be achieved based on the differentiated service. The situation classification is based on the type of use cases that can exist within a tenant in a micro-operator network and it can be defined as follows:

- **Situation 1** describes when one slice instance is allocated for one tenant. This situation can be applicable when a tenant's use case requires a specific network service, such as an eMBB, uRLLC or mMTC services. Thus, only the required service slice will be distributed across the tenant.

- **Situation 2** describes when one slice instance is to be distributed or allocated across multiple tenants. This situation can happen due to shared subnet within the network which is now reflected outward to the tenants. Thus, one slice communication service will be shared amidst multiple tenants requiring the same network service.

- **Situation 3** describes when a tenant's use case requires more than one slice instance. This happens when a use case requires multiple differentiated services based on the network requirements. The use cases within a tenant can require an eMBB and uRLLC slice for high throughput and low latency. As such the tenant's network requirement will need different slices allocated to a single use case

- **Situation 4** describes a multi-tenant case where different slice instances are allocated for multiple tenants' use cases. Unlike situation 3 where different slices are allocated to a single tenant's use case, in this situation, multiple slice services are allocated to multiple tenants.

The situation types described can be used to define how slice allocation will happen for each deployment scenario of a micro-operator network or any other possible use cases such as for UAVs, industrial robots, etc. Considering different micro-operator network deployment scenarios, Table I highlights a mapped relation between each deployment scenario of a micro-operator and the multitenant-slice situation type that can exist.

To describe the network slicing communication service distribution for different multitenant-slice situation types in a micro-operator network, we used a high-level network slicing architecture (Fig. 3). The slicing architecture in Fig. 3 divides the end-to-end slicing implementation of a micro-operator into four layers.

TABLE 1. MICRO-OPERATOR DEPLOYMENT SCENARIOS MAPPED WITH SITUATION TYPES

| Micro-operator | Deployment Scenarios | Deployment Description | Possible Situation types |
|---|---|---|---|
| Closed | Deployment A | A closed network is responsible for one tenant at a single location. | Situation 1 where one slice service is allocated to a single tenant. |
| Closed | Deployment B | A closed network is responsible for a set of tenants at different locations. | Situation 1 or Situation 2 due to single or multiple tenants at different locations. |
| Open | MNO Open | Targeted at MNO subscribers within a given locality based on the service agreement with the MNO. The micro-operator may be responsible for subscribers from one MNO or from multiple MNOs. | Situation 2 or Situation 4. Situation 2 for subscribers from one MNO network sharing a single tenant slice and situation 4 for subscribers from multiple MNO networks sharing multiple slices. |
| Open | Public Open | Targeted at the general public. | Situation 2 where one slice service is reserved for public use. |
| Mixed | Option A | Micro-operator needs services from the MNO such as wide area access, remote monitoring, etc. thereby configuring tenant's slices with MNO resources. | Situation 3 where a micro-operator service needs NSSIs from both MNO and the micro-operator. |
| Mixed | Option B | MNO needs services from the micro-operator network such as better service for its subscribers within the micro-operator's network and using the tenant's broadband slice for extending indoor coverage. | Situation 3 or Situation 4. Situation 3 for a communication service formed for both micro-operator NSI and MNO NSI. Situation 4 for multiple NSIs serving multiple tenants including MNO tenants. |

**The service layer** (for OSS/BSS and policy control), **the Slicing MANO layer** (to cover NSMF and NSSMF implementation using OSM), **the Resource layer** (for virtual resource allocation, management and orchestration using ETSI NFV), and **the Multi-tenant layer** (for implementing CSMF slice request aggregation and slice instance distribution). However, in this paper, we focused on the multi-tenant layer of the architecture which determines the implementation of the CSMF and how the communication service for every network slice instance is being distributed across different tenants. Thus, the multi-tenant layer is in charge for handling multi-tenancy within the micro-operator network.

Unlike the previous architectures such as in [13], [17] that described multi-tenancy only in terms of how slice requests will be handled by a slice broker, our proposed architecture makes a clear description on how to achieve multi-tenancy by extending the 3GPP CSMF capabilities to implement network slice communication services distribution across different tenants, and subsequently their end users. From Fig.3, we present an implementation technique for the CSMF capability with a multi-tenant manager and a communication service orchestrator. The former is responsible for slice request handling and aggregation from each tenant, and the latter is responsible for the communication service distribution to the required tenants. The reason for implementing the CSMF function with the multi-tenant manager and the communication service orchestrator in Fig.3, is to be able to separately achieve all possible types of slice requests from every deployment scenario and to manage all possible situation types that can exist during slice instance distribution. This approach will efficiently implement the CSMF.

*A. The Multi-Tenant Manager*

The multi-tenant manager is responsible for the aggregation of different slice requests from individual tenants and forwarding those requests to the service layers for policy, charging and operation support. Similar to a slice broker, every tenant request with a slice request ID is abstracted to the multi-tenant manager and then forwarded through the service layer to the slicing MANO layer where the NST will be implemented and the slice instance will be created across the AN and CN. The slice request ID will be dependent on the tenant's use case, end users, deployment scenario, and the situation type to determine the needed network resources in forming the slice. The slice request will also contain tenant information such as deployment type, location of tenants, network requirement, shared constituents and confirmation if an external resource will be required or not. All these parameters will be abstracted to the slicing MANO layer for the slice instance creation. Then, after the slice has been instantiated, the communication service orchestrator will be responsible for intelligently identifying each tenant's slice and the allocation of the tenants' slice to the respective end users.

*B. The Communication Service Orchestrator*

The communication service orchestrator extends the full functionalities of the CSMF where it is responsible for the transmission of the "per tenant" instantiated slice i.e. (end- to-end NSI), as an established communication service back to the tenants. The communication service orchestrator will implement this by mapping back the tenant request ID to a slice creation ID. The communication service orchestrator will also be responsible for combining multiple NSIs, i.e. slices instances for differentiated services and transmitting them to their respective tenant(s). In order to approach the different situation types that can exist when a slice is to be allocated to multiple tenants, the communication service orchestrator will be implemented in four layers. The layered approach proposed in this paper is similar to the approach used by OSM [12] in the NSI creation described in earlier section. In the proposed technique, at the communication service orchestrator, the basic layer will be an instantiated NSI. A single NSI will be implemented for the basic situation type that can exist (e.g. situation 1 when one network slice is attributed to a single tenant).

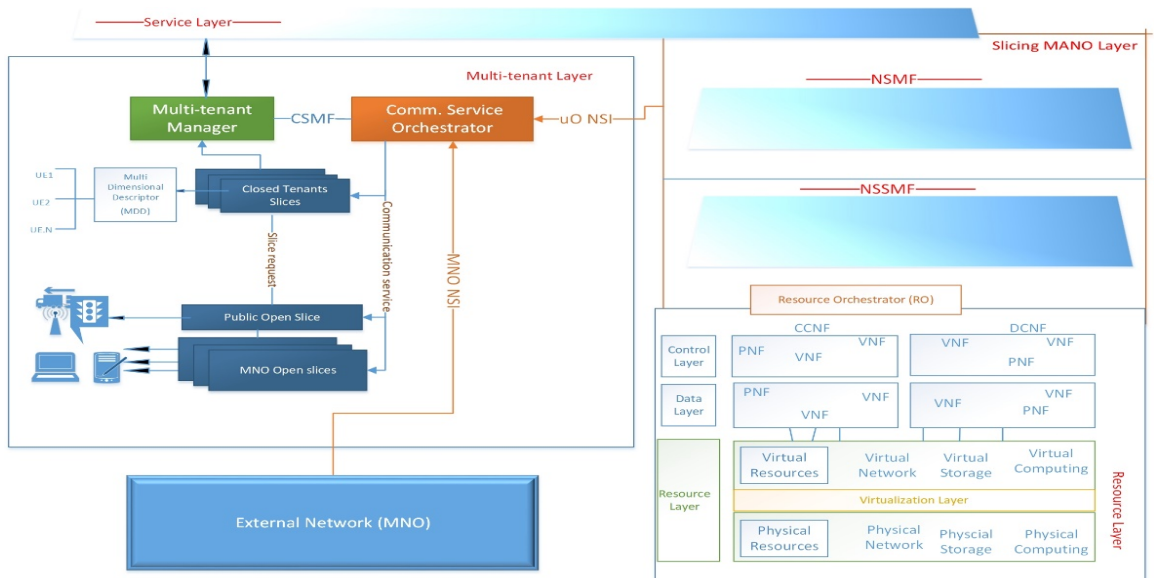

Fig 3. Micro-operator slicing architecture with focus on the multi-tenant layer.

As depicted in Fig. 4, each layer will represent each situation in a hierarchical form; Layer 1 will implement situation 1, layer 2 will implement situation 2, depending on layer 1, layer 3 will implement situation 3, and layer 4 will implement situation 4 depending on layer 2 and layer 3. As shown in Fig. 4, each layer is dependent on other layers, this implies that situations 2, 3, or 4 can only be implemented if there is a layer 1 (i.e., an instantiated NSI). Therefore, the implementation of each layer to achieve the possible situation types can be described as follows:

- **Layer 1:** This will serve as the basic layer, and it will be NSI specific for each tenant. The NSI is allocated to each of the end users of a single tenant. A layer 1 template will be created containing parameters such as the slice creation ID in relation to the slice request ID, the situation types to determine if the slice will be for a single slice. Hence, if the slice is requested for one tenant, Layer 1 is activated and, the slice instance communication service will be allocated to the tenant.

- **Layer 2:** Layer 2 implementation for multitenant-slice situation type 2 will be achieved with a template containing a layer 1 description. However, in this case, a higher description layer will be implemented. This means that a Layer 2 template will contain a layer 1 description, but a layer 1 description will not contain a layer 2 description as seen in figure 4. For layer 2, a service broker will be introduced to manage the network resource of a single network slice instance amidst different tenants. The implementation of the broker will be performed in an action file (an executable file) that will be described using a NS primitive in Juju charm (a solution for automating action performed in a virtualized network). In the layer 2 description, the number of tenants will be increased for a single network and hence the service broker action will be implemented.

- **Layer 3:** Layer 3 implementation for multitenant-slice situation type 3 involves a single tenant being allocated multiple slice instances, whereby each instance is responsible for a specific network service. The implementation of layer 3 will be very vital to achieve the full capabilities of the CSMF and many use cases require this situation outside a micro-operator such as UAV use cases that require eMBB and uRLLC slices. The layer 3 implementation will be achieved by combining multiple layer 1 templates and using a service broker to achieve the combination of two slices for a single tenant. The service broker will also be implemented in an action file described with a Juju charm, which will implement the allocation of both slices to a single tenant. Furthermore, since all layer 1 templates are representing different NSIs identified for a network service, the layer 3 template will specify the required slice instances (i.e., required layer 1 templates) that will be combined. This will be based on the slice creation ID. Finally, the service broker can also combine multiple NSIs from a single operator or multiple NSIs from multiple operators. The implementation of layer 3 to achieve multi-tenant slice situation type 3 is rather vital to a full CSMF implementation.

- **Layer 4:** Layer 4 implementation for situation type 4 is based on multiple layer 2 (and even layer 3) templates, as the service broker will implement the rest. If a layer 2 template is used, the service broker will combine multiple layer 2 templates and call an action file to combine multiple NSIs. There will be multiple slice instances services implemented for multiple tenants. However, if a layer 3 template is used, it will be easier as only the number of tenants will be increased.

Since the layered approach presented in this paper is hierarchical, layer 4 can only be implemented with layer 2 or layer 3 templates. In the same way, layer 2 or layer 3 can only be implemented with a layer 1 template. As such, layer 4 is indirectly dependent to layer 1. In general, our proposed solution will cover every single future use case that can exist in terms of network slice communication service distribution across multiple tenants.

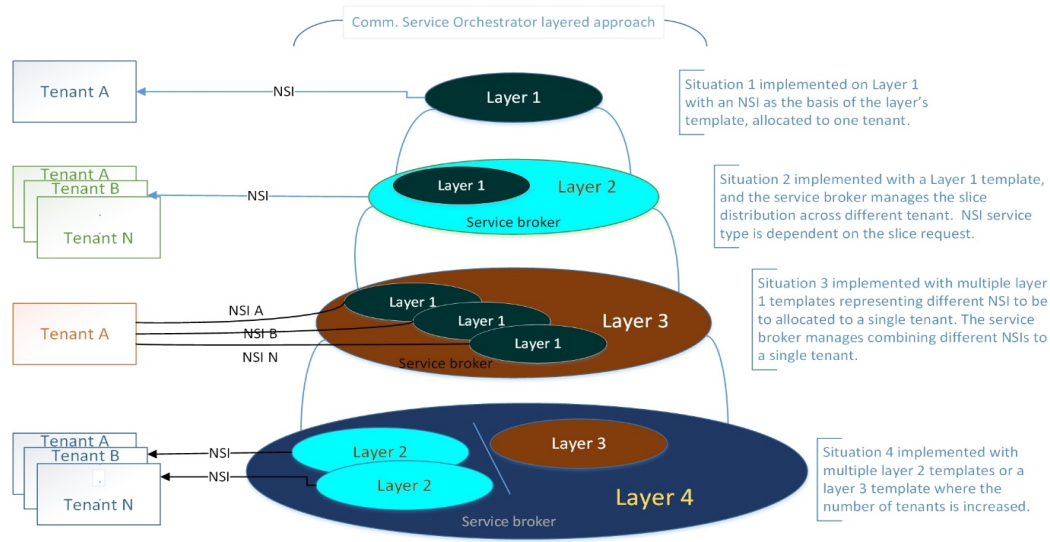

Fig 4. Communication service orchestrator implementation of multitenant-slice situation types.

## V. CONCLUSION

In this paper, an implementation technique is presented to address how the network slice communication service will be distributed across different tenants of a micro-operator network. This paper proposed the expansion of the CSMF to accommodate a multi-tenant manager and the communication service orchestrator. The paper further detailed the multitenant-slice situation types that can exist, and how the communication service orchestration will implement the situation types based on a layered approach. The proposed framework presented in this paper will serve as a generic architecture for future research, integrating the proposed CSMF capabilities with the existing orchestration implementations. Future research within this area will attempt to complete the implementation of the CSMF, targeted to accommodate other use cases starting with UAV scenarios.

## ACKNOWLEDGMENT

This work was partially supported by the European Union's Horizon 2020 Research and Innovation Program under the 5G!Drones project (Grant No. 857031), and by the Academy of Finland 6Genesis project (Grant No. 318927).